\documentclass[aps,prd,preprint,groupedaddress,preprintnumbers,longbibliography,nofootinbib]{revtex4-1}
\newcommand{\ep}{\epsilon}
\newcommand{\nn}{\nonumber}
\newcommand{\lb}{\left\lbrace}
\newcommand{\rb}{\right\rbrace}
\newcommand{\vev}[1]{\left\langle #1 \right\rangle}
\newcommand{\vvev}[1]{\left\langle\kern-0.3em\left\langle #1
    \right\rangle\kern-0.3em\right\rangle} 
\newcommand{\Op}{\mathcal{O}}
\newcommand{\E}{\mathcal{E}}
\usepackage{amsmath}

\begin{document}

\preprint{KOBE-TH-17-01}

\title{Conformal invariance for Wilson actions}


\author{H.~Sonoda}\email[]{hsonoda@kobe-u.ac.jp}\affiliation{Physics
  Department, Kobe University, Kobe 657-8501, Japan}
%\thanks{}
%\affiliation{}


\date{\today}

\begin{abstract}
  We discuss the realization of conformal invariance for Wilson
  actions using the formalism of the exact renormalization group.
  This subject has been studied extensively in the recent works of
  O.~J.~Rosten.  The main purpose of this paper is to reformulate
  Rosten's formulas for conformal transformations using a method
  developed earlier for the realization of any continuous symmetry in
  the exact renormalization group formalism.  The merit of the
  reformulation is simplicity and transparency via the consistent use
  of equation-of-motion operators.  We derive equations that imply the
  invariance of the Wilson action under infinitesimal conformal
  transformations which are non-linearly realized but form a closed
  conformal algebra.  The best effort has been made to make the paper
  self-contained; ample background on the formalism is provided.
\end{abstract}

%\keywords{}

\maketitle

\section{Introduction\label{section-introduction}}

The study of conformally invariant field theories (in dimensions
$D>2$) was initiated long ago by J.~Wess \cite{Wess:1960} with a hope
that conformal invariance constrains a theory more than scale
invariance, since the latter is implied by the former.  Requirement of
conformal invariance seemed much stronger than that of scale
invariance at first sight, but the difference turned out to be subtle.
In the seminal work \cite{Polchinski:1987dy}, J.~Polchinski showed the
equivalence of conformal invariance to the vanishing of the trace of
the energy-momentum tensor; scale invariance requires the vanishing of
only its integral.  The question of whether scale invariance implies
conformal invariance has attracted much attention lately, and we would
like to refer the reader to a recent review by Y.~Nakayama
\cite{Nakayama:2013is} and references therein.

The subject of this paper is realization of conformal symmetry using
Wilson actions.\cite{Wilson:1973jj} This was recently taken up by
O.~J.~Rosten \cite{Rosten:2014oja} and also by Delamotte, Tissier, and
Wschebor \cite{Delamotte:2015aaa}.  Rosten has extended his work
further in \cite{Rosten:2016nmc, Rosten:2016zap}.  It is the recent
works of Rosten (especially \cite{Rosten:2014oja} and
\cite{Rosten:2016zap}) that we wish to improve upon by using the
method of symmetry realization developed and reviewed in
\cite{Igarashi:2009tj}.  We aim to add simplicity and transparency to
the structure of conformal transformations in the exact
renormalization group formalism.

Wilson actions come with a finite momentum cutoff, and it is generally
accepted that only the physics at scale below the cutoff is
effectively described by Wilson actions.  This is indeed the case with
a generic Wilson action, but there are exceptions.  Those Wilson
actions flowing out of a fixed point under the renormalization group
transformations correspond to a continuum limit, and the physics at
all momentum scales are described by the Wilson actions.  (In
\cite{Wilson:1973jj} these Wilson actions form a finite dimensional
space $S(\infty)$.)  Hence, if the continuum limit of a theory has
symmetry, we can realize the symmetry using its Wilson action.  Now, a
fixed point of the renormalization group transformation is a continuum
limit.  If the limit possesses conformal symmetry, its Wilson action
must realize the symmetry, too.

The method of \cite{Igarashi:2009tj} has recently been applied to the
construction of the energy-momentum tensor in \cite{Sonoda:2015pva}.
Our expression of special conformal transformation
(\ref{SigmaK-definition}) was in fact first derived there from the
assumption of the vanishing trace.  We summarize this derivation in
Appendix \ref{appendix-EM}.

We organize the paper as follows.  In Sect.~\ref{section-algebra} we
introduce infinitesimal conformal transformations of the elementary
scalar field in $D$-dimensional Euclidean space.  In
Sect.~\ref{section-invariance}, we review quickly how to express
continuous symmetry of a Wilson action in terms of equation-of-motion
composite operators.  Then, in
Sect.~\ref{section-conformalinvariance}, we construct
equation-of-motion composite operators for the conformal symmetry, and
subsequently in Sect.~\ref{section-realization} we construct the
products of the infinitesimal transformations to show the closure of
the algebra.  Sects.~\ref{section-conformalinvariance} and
\ref{section-realization} constitute the main part of this paper.  In
Sect.~\ref{section-invariance-WGamma} we rewrite the invariance of the
Wilson action as that of the associated generating functional and 1PI
action.  In Sect.~\ref{section-example} we construct the 1PI action of
a Wilson-Fisher fixed point in $D=4-\ep$ dimensions to first order in
$\ep$.  We extend the conformal transformation to the scalar composite
operators in Sect.~\ref{section-composite} before we conclude the
paper in Sect.~\ref{conclusion}.

We have kept the main text reasonably short by relegating the
technicalities to five appendices.  The effort has been made to make
this technical paper an easy read; the first reading of the main text
had better be done without referring to the appendices.  We have
adopted the following notation
\begin{equation}
\int_p = \int \frac{d^D p}{(2 \pi)^D},\quad
\delta (p) = (2 \pi)^D \delta^{(D)} (p),\quad p \cdot q = p_\mu q_\mu =
\sum_{\mu=1}^D p_\mu q_\mu
\end{equation}
to simplify the formulas.

\section{Conformal algebra\label{section-algebra}}

We consider a real scalar field theory in $D$ dimensional Euclidean
space.  We first consider the field in coordinate space.
Infinitesimal conformal transformations act on the field as follows
\cite{Wess:1960}:
\begin{subequations}
\begin{eqnarray}
D_\mu^T \phi (x) &\equiv& \frac{1}{i} \partial_\mu \phi (x),\\
D^R_{\mu\nu} \phi (x) &\equiv& \left(x_\mu \partial_\nu -
  x_\nu \partial_\mu\right) \phi (x),\\
D^S \phi (x) &\equiv& \left(x_\mu \partial_\mu + \frac{D-2}{2} +
  \gamma\right) \phi (x),\\
D^K_\mu \phi (x) &\equiv& \frac{1}{i} \left( x_\mu x_\nu \partial_\nu - \frac{1}{2} x^2
  \partial_\mu + \left(\frac{D-2}{2} + \gamma \right) x_\mu
\right) \phi (x),
\end{eqnarray}
\end{subequations}
where $\frac{D-2}{2} + \gamma$ is the full scale dimension of the
scalar field including the anomalous dimension $\gamma$.  We have
chosen the superscript $T$ for translation, $R$ for rotation, $S$ for
scale transformation, and $K$ for the special conformal transformation that
results from the succession of inversion, translation, and inversion.
The algebra of the differential operators is closed, and is called the
conformal algebra \cite{Wess:1960}:
\begin{subequations}
\label{algebra}
\begin{eqnarray}
\left[ D^T_\mu, D^T_\nu \right] &=& 0,\\
\left[ D^R_{\alpha\beta}, D^R_{\gamma\delta}\right] &=&
\delta_{\beta\gamma} D^R_{\alpha\delta} - \delta_{\beta\delta}
D^R_{\alpha\gamma} - \delta_{\alpha\gamma} D^R_{\beta\delta} +
\delta_{\alpha\delta} D^R_{\beta\gamma},\\
\left[D^R_{\mu\nu}, D^T_\alpha\right] &=& - \delta_{\mu\alpha} D_\nu^T +
\delta_{\nu\alpha} D_\mu^T,\\
\left[D^S, D^T_\mu \right] &=& D_\mu^T,\\
\left[D^S, D^R_{\mu\nu} \right] &=& 0,\\
\left[D^K_\mu, D^K_\nu\right] &=& 0,\\
\left[D^K_\mu, D^T_\nu\right] &=&  D^S \delta_{\mu\nu} + 
D_{\mu\nu}^R,\\
\left[D^K_\mu, D^R_{\alpha\beta}\right] &=& \delta_{\mu\alpha}D^K_\beta
- \delta_{\mu\beta} D^K_\alpha,\\
\left[ D^K_\mu ~,~ D^S \right] &=& - D^K_\mu .
\end{eqnarray}
\end{subequations}

We formulate the Wilson action in momentum space; it is more
convenient to rewrite the above transformations in momentum space.
Denoting the Fourier transform of the scalar field by
\begin{equation}
\phi (p) \equiv \int d^D x\, e^{- i p x} \phi (x)\,,
\end{equation}
we obtain
\begin{subequations}
\label{D-definition}
\begin{eqnarray}
D_\mu^T (p) \phi (p) &=& p_\mu \phi (p)\,,\\
D^R_{\mu\nu} (p) \phi (p) &=& \left( p_\mu \frac{\partial}{\partial
                            p_\nu} - p_\nu \frac{\partial}{\partial
                            p_\mu}\right) \phi (p)\,,\\
D^S (p) \phi (p) &=& \left( - p_\mu \frac{\partial}{\partial p_\mu} -
                   \frac{D+2}{2} + \gamma \right) \phi (p)\,,\\
D^K_\mu (p) \phi (p) &=& \left( -  p_\nu\frac{\partial^2}{\partial
                       p_\mu \partial p_\nu} +  \frac{1}{2} p_\mu
                       \frac{\partial^2}{\partial p_\nu \partial
                       p_\nu} + \left(- \frac{D+2}{2} + \gamma\right)
                       \frac{\partial}{\partial p_\mu} \right) \phi (p)\,.
\end{eqnarray}
\end{subequations}
The above $D (p)$'s obey the same conformal algebra as
(\ref{algebra}): for example, we obtain
\begin{equation}
\left[ D^K_\mu (p), D^T_\nu (p) \right] =  D^S (p) \delta_{\mu\nu} +
 D^R_{\mu\nu} (p)\,.
\end{equation}

\section{Invariance of a Wilson action\label{section-invariance}}

The infinitesimal conformal transformations are linear transformations
of the scalar field.  There is no guarantee, however, that they are
realized as linear transformations for the Wilson action.  Suppose
that the Wilson action $S[\phi]$ is ``invariant'' under an
infinitesimal transformation $\Delta \phi (p)$ of the field variable
$\phi (p)$.  Since the exponentiated Wilson action $e^S$ is the
measure of functional integration, the invariance of the theory under
the infinitesimal transformation amounts to
\begin{equation}
\int_p \left( \Delta \phi (p) \frac{\delta S}{\delta \phi (p)} +
  \frac{\delta}{\delta \phi (p)} \Delta \phi (p) \right)  = 0\,,
\end{equation}
where the second term comes from the Jacobian.  This can be written as
\begin{equation}
\int_p \frac{\delta}{\delta \phi (p)} \left(\Delta \phi (p)\,
  e^S \right) = 0\,.
\end{equation}

In the ERG formalism we choose
\begin{equation}
\Delta \phi (p) = K(p) \Op (p)\,.
\end{equation}
$K(p)$ is a positive momentum cutoff function: it depends only
on $p^2$, is nearly $1$ for momenta low compared with the cutoff
$p=1$, and decreases rapidly for $p \gg 1$.  $\Op (p)$ is a composite
operator (i.e., a functional of $\phi$) with momentum $p$.  Using the
above $\Delta \phi$, we obtain the invariance as
\begin{equation}
\int_p K(p) \frac{\delta}{\delta \phi (p)} \left( \Op (p)\,e^S
\right) = 0\,.
\end{equation}
This is the general form of the equation of motion in the ERG
formalism.  The equation of motion implies the Ward-Takahashi identity
for the correlation functions:
\begin{equation}
\sum_{i=1}^n \vvev{\phi (p_1) \cdots \Op (p_i) \cdots \phi (p_n)} = 0\,,
\end{equation}
where the $i$-th $\phi$ is replaced by $\Op$.  Note that we use
$\vvev{\cdots}$ for the continuum limit of correlation functions.  We
refer the reader to Appendices \ref{appendix-ERG} \&
\ref{appendix-WGamma}, where we give technical details on the ERG
formalism such as modified correlation functions and
equations-of-motion composite operators.

\section{Conformal invariance\label{section-conformalinvariance}}

For infinitesimal conformal transformations, we choose $\Op (p)$ of
the previous section as $D (p) \Phi (p)$, where $D (p)$ is one of
$D^T_\mu (p), D^R_{\mu\nu} (p), D^S (p), D^K_\mu (p)$ introduced in
sect.~\ref{section-algebra}.  $\Phi (p)$ is a composite operator
defined by
\begin{equation}
\Phi (p) \equiv e^{-S} \frac{1}{K(p)} \left(\phi (p) +
  \frac{k(p)}{p^2} \frac{\delta}{\delta \phi (-p)} \right) e^S\,,\label{Phi-def}
\end{equation}
where $S$ is the Wilson action, and $K, k$ are cutoff functions.
$\Phi (p)$ has the same correlation functions as the elementary field
$\phi (p)$:
\begin{equation}
\vvev{\Phi (p) \phi (p_1) \cdots \phi (p_n)} = \vvev{\phi (p) \phi
  (p_1) \cdots \phi (p_n)}\,.
\end{equation}
See Appendix \ref{appendix-ERG} for the precise definition of both
sides.  Hence,
\begin{eqnarray}
\vvev{D (p) \Phi (p) \phi (p_1) \cdots \phi (p_n)} &\equiv& D(p) \vvev{\Phi
  (p) \phi (p_1) \cdots \phi (p_n)} \nn\\
&=& D(p) \vvev{\phi (p) \phi (p_1)  \cdots \phi (p_n)}\,.
\end{eqnarray}

We now introduce the following the equation-of-motion composite operators:
\begin{subequations}
\label{Sigma-definition}
\begin{eqnarray}
\Sigma_\mu^T &\equiv& - e^{-S} \int_p K(p) \frac{\delta}{\delta \phi (p)} \left(
  D^T_\mu (p) \Phi (p) \,e^S \right)\,,\\
\Sigma^R_{\mu\nu} &\equiv&  - e^{-S} \int_p K(p) \frac{\delta}{\delta
  \phi (p)} \left( 
  D^R_{\mu\nu} (p) \Phi (p) \,e^S \right)\,,\\
\Sigma^S &\equiv&  - e^{-S} \int_p K(p) \frac{\delta}{\delta \phi (p)} \left(
  D^S (p) \Phi (p) \,e^S \right)\,,\\
\Sigma^K_\mu &\equiv&  - e^{-S} \int_p K(p) \frac{\delta}{\delta \phi (p)} \left(
  D^K_\mu (p) \Phi (p) \,e^S \right)\,.\label{SigmaK-definition}
\end{eqnarray}
\end{subequations}
These carry no momentum.
The conformal invariance amounts to the vanishing of the above operators:
\begin{equation}
  \Sigma_\mu^T = \Sigma^R_{\mu\nu} = \Sigma^S = \Sigma^K_\mu =
  0\,.\label{conformal-invariance} 
\end{equation}
Substituting these into the correlation functions, we obtain the
following Ward-Takahashi identities:
\begin{subequations}
\begin{eqnarray}
\sum_{i=1}^n D_\mu^T (p_i) \vvev{\phi (p_1) \cdots \phi (p_n)} &=& 0,\\
\sum_{i=1}^n  D_{\mu\nu}^R (p_i)  \vvev{\phi
  (p_1) \cdots \phi (p_n)} &=& 0,\\
\sum_{i=1}^n D^S (p_i) \vvev{\phi (p_1)
  \cdots \phi (p_n)} &=& 0,\\
\sum_{i=1}^n D^K_\mu (p_i) \vvev{\phi (p_1) \cdots \phi (p_n)}&=&0.
\end{eqnarray}
\end{subequations}

Note that the scale invariance, given by $\Sigma^S = 0$, is nothing
but the ERG differential equation for a fixed point Wilson action,
which is usually given in the form \cite{Wilson:1973jj}
\begin{eqnarray}
  0 &=& \int_p \left( - p \cdot \partial_p \ln K(p) + \frac{D+2}{2} -
    \gamma + p \cdot \partial_p \right) \phi (p) \cdot
  \frac{\delta}{\delta \phi (p)} \,e^S\nn\\
  && + \int_p \left( p \cdot \partial_p \ln \frac{k(p)}{K(p)^2} - 2
    \gamma \right) \frac{k(p)}{p^2} \frac{1}{2} \frac{\delta^2}{\delta
    \phi (p) \delta \phi (-p)} \,e^S\,.\label{ERGequation}
\end{eqnarray}
This rewriting has been explained in Appendix B of \cite{Sonoda:2015pva}.

As for the special conformal invariance $\Sigma^K_\mu = 0$, an
equivalent formula was first derived by Rosten as (3.25) in
\cite{Rosten:2014oja}.  The particular form $\Sigma_\mu^K$ given by
(\ref{SigmaK-definition}) was first obtained in
\cite{Sonoda:2015pva}. (This is briefly explained in Appendix
\ref{appendix-EM}.)  Rosten has rewritten his result as (2.79b) in
\cite{Rosten:2016nmc}.  We will explain how to derive (a formula
similar to) his (2.79b) by rewriting our $\Sigma^K_\mu = 0$ in
Appendix \ref{appendix-SigmaK}.

\section{Realization of the conformal algebra\label{section-realization}}

So far we have only discussed the invariance of the Wilson action
under infinitesimal conformal transformations.  The conformal
transformations form a closed algebra, and the algebraic structure
must be realized on the Wilson action.

For realization of the algebra, we need the product of two
infinitesimal transformations.  Let $D_i\,(i=1,2)$ be two of the
infinitesimal transformations $D^T_\mu, D^R_{\mu\nu}, D^S, D^K_\mu$,
and we denote
\begin{equation}
\Sigma_i \equiv - e^{-S} \int_p K(p) \frac{\delta}{\delta \phi (p)}
\left( D_i (p) \Phi (p) \,e^S \right)\,.
\end{equation}
We construct the product as
\begin{equation}
\Sigma_1 * \Sigma_2 \equiv -
e^{-S} \int_p K(p) \frac{\delta}{\delta \phi (p)} \lb 
    D_1 (p)\left[ \Phi (p) \Sigma_2 \right] \, e^S\rb\,,
\end{equation}
where 
\begin{equation}
\left[\Phi (p) \Sigma_2 \right] \equiv \Phi (p) \Sigma_2 +
\frac{k(p)}{p^2 K(p)} \frac{\delta  \Sigma_2}{\delta \phi (-p)}
\end{equation}
is a composite operator corresponding to the product of $\Phi (p)$ and
$\Sigma_2$.  (See Appendix A.)  Using
\begin{equation}
  \vvev{\left[ \Phi (p_i) \Sigma_2 \right] \phi (p_1) \cdots
    \widehat{\phi (p_i)} \cdots \phi (p_n)}
  = \vvev{\Sigma_2 \, \phi (p_1) \cdots \phi (p_n)}
\end{equation}
(where the hat above $\phi (p_i)$ implies omission), 
we obtain
\begin{eqnarray}
&&\vvev{\Sigma_1 * \Sigma_2\, \phi (p_1) \cdots \phi (p_n)}\nn\\
&&= \sum_{i=1}^n D_1 (p_i) \vvev{\phi (p_1) \cdots \left[ \Phi (p_i)
    \Sigma_2 \right] \cdots \phi (p_n)}\nn\\
&&= \sum_{i=1}^n D_1 (p_i) \vvev{\Sigma_2\, \phi (p_1) \cdots 
    \phi (p_n)}\nn\\
&&= \sum_{i=1}^n D_1 (p_i) \sum_{j=1}^n D_2 (p_j) \vvev{\phi (p_1)
  \cdots \phi (p_n)}\nn\\
&&= \sum_{i=1}^n \left( D_1 (p_i) D_2 (p_i) + \sum_{j\ne i} D_1 (p_i)
  D_2 (p_j) \right) \vvev{\phi (p_1) \cdots \phi (p_n)}\,.
\end{eqnarray}
Therefore, we obtain
\begin{equation}
\vvev{\left(\Sigma_1 * \Sigma_2 - \Sigma_2 * \Sigma_1 \right) \, \phi
  (p_1) \cdots \phi (p_n)} = \sum_{i=1}^n [D_1 (p_i), D_2 (p_i)]
\,\vvev{\phi (p_1) \cdots \phi (p_n)}\,.
\end{equation}
This implies
\begin{equation}
\Sigma_1 * \Sigma_2 - \Sigma_2 * \Sigma_1 =
- e^{-S} \int_p K(p) \frac{\delta}{\delta \phi (p)} \left( \left[D_1 (p),
    D_2 (p)\right] \Phi (p)\, e^S \right)\,.
\end{equation}
Hence, the algebra of $D$'s translates into the algebra of $\Sigma$'s.

The higher products of $\Sigma$'s can be defined recursively as
\begin{equation}
\Sigma_1 * \Sigma_2 * \cdots * \Sigma_I
\equiv - \int_p K(p) \frac{\delta}{\delta \phi (p)} \lb D_1 (p) \left[
  \Phi (p) \Sigma_2 * \cdots * \Sigma_I \right] \,e^S\rb\,,
\end{equation}
so that
\begin{eqnarray}
&&\vvev{\Sigma_1 * \Sigma_2 * \cdots * \Sigma_I \,\phi (p_1) \cdots \phi
  (p_n)} \nn\\
&&= \sum_{i_1 = 1}^n D_1 (p_{i_1}) \sum_{i_2=1}^n D_2 (p_{i_2})
\cdots \sum_{i_I = 1}^n D_I (p_{i_I})\, \vvev{\phi (p_1) \cdots \phi
  (p_n)}\,.
\end{eqnarray}

\section{Conformal symmetry for the generating functional and 1PI
  action\label{section-invariance-WGamma}}

We wish to rewrite the equation-of-motion composite operators
(\ref{Sigma-definition}) in terms of the generating functional $W[J]$
of connected correlations and 1PI action $\Gamma [\Phi]$ associated
with the Wilson action $S[\phi]$.  The Wilson action results from the
integration of the field with momenta above $p = 1$; the field with
momenta below $p=1$ has not been integrated for the generating
functional $W [J]$ and 1PI action $\Gamma [\Phi]$.  $W[J]$ and
$\Gamma [\Phi]$ depend only on a particular combination of the two
cutoff functions
\begin{equation}
R(p) \equiv \frac{p^2}{k(p)} K(p)^2
\end{equation}
which is non-vanishing (if not divergent) at $p=0$, and decreases
rapidly for $p \gg 1$.  ($R(p)$ is often called a scale dependent
squared mass in the ERG literature.)

Formulas necessary for rewriting conformal invariance of $S[\phi]$ as
that of $W[J]$ have been summarized in Appendix B:
\begin{equation}
J(p) \equiv \frac{R(p)}{K(p)} \phi (p),\quad
\frac{\delta W[J]}{\delta J(-p)} = \Phi (p),\quad
W[J] = S[\phi] + \frac{1}{2} \int_p \frac{1}{R(p)} J(p) J(-p)\,.
\end{equation}
Similarly, the formulas
\begin{equation}
\lb\begin{array}{c}
J(p) = - R(p) \Phi (p) + \frac{\delta \Gamma [\Phi]}{\delta \Phi
  (-p)},\quad
\frac{\delta^2 W[J]}{\delta J(p) \delta J(q)} = G_{p,q} [\Phi]\,,\\
\Gamma [\Phi] - \frac{1}{2} \int_p R(p) \Phi (p) \Phi (-p) = W[J] -
\int_p J(p) \Phi (-p)\,,
\end{array}\right.
\end{equation}
which are necessary for rewriting the conformal invariance of $W [J]$
as that of $\Gamma [\Phi]$, are also summarized in Appendix B.
Assuming the rotational invariance of $R(p)$ and $R(-p) = R(p)$, we
obtain the following results:
\begin{enumerate}
\item $T$ (translation invariance)
\begin{subequations}
\begin{eqnarray}
\int_p D_\mu^T (p) J(p) \cdot \frac{\delta W[J]}{\delta J(p)} &=& 0\,,\\
\int_p D_\mu^T (p)  \Phi (p) \cdot \frac{\delta \Gamma [\Phi]}{\delta
  \Phi (p)} &=& 0\,. 
\end{eqnarray}
\end{subequations}
\item $R$ (rotation invariance)
\begin{subequations}
\begin{eqnarray}
    \int_p D^R_{\mu\nu} (p) J(p) \cdot \frac{\delta
      W[J]}{\delta J(p)} &=& 0\,,\\
    \int_p D^R_{\mu\nu} (p) \Phi (p) \cdot
    \frac{\delta \Gamma [\Phi]}{\delta \Phi (p)} &=& 0\,.
\end{eqnarray}
\end{subequations}
\item $S$ (scale invariance)
\begin{subequations}
\label{Sinvariance-WGamma}
\begin{eqnarray}
&& \int_p J(-p) D^S (p)
\frac{\delta W[J]}{\delta J(-p)}\nn\\
&& \quad + \int_p \left(- p \cdot \partial_p + 2 - 2\gamma \right) R(p)
\cdot \frac{1}{2} \lb \frac{\delta^2 W[J]}{\delta J (p)
  \delta J(-p)} + \frac{\delta W[J]}{\delta J(p)} \frac{\delta
  W[J]}{\delta J(-p)} \rb = 0\,,\\
&&- \int_p \frac{\delta \Gamma [\Phi]}{\delta \Phi (p)} D^S (p) \Phi (p)
 + \int_p \left(- p \cdot \partial_p + 2 - 2\gamma \right) R(p)
\cdot \frac{1}{2} G_{p, -p} [\Phi] = 0\,,
\end{eqnarray}
\end{subequations}
where the integrals with $R$ have been simplified by partial
integration.
\item $K$ (special conformal invariance)
\begin{subequations}
\label{Kinvariance-WGamma}
\begin{eqnarray}
&&\int_p J(-p) D^K_\mu (p) \frac{\delta W[J]}{\delta J(-p)}
\label{Kinvariance-W}\\
&&\quad + \frac{1}{2} \int_p \left( - p \cdot \partial_p + 2 -
  2\gamma \right) R(p)
\cdot \frac{\partial}{\partial p_\mu} \lb
\frac{\delta^2 W[J]}{\delta J (p) \delta J(-q)} 
+ \frac{\delta W[J]}{\delta J(p)} \frac{\delta
  W[J]}{\delta J(-q)} \rb\Big|_{q=p} = 0\,,\nn\\
&&- \int_p \frac{\delta \Gamma [\Phi]}{\delta \Phi (p)} D^K_\mu (p) \Phi
(p) + \frac{1}{2} \int_p  \left( - p \cdot \partial_p + 2 -
  2\gamma \right) R(p) \cdot \frac{\partial G_{-p,q}
[\Phi]} {\partial p_\mu} \Big|_{q=p} = 0\,,\label{Kinvariance-Gamma}
\end{eqnarray}
\end{subequations}
where $q$ is set equal to $p$ only after the derivative is taken.  The
integrals with $R$ have been simplified by partial integration.  This
step is explained in Appendix \ref{appendix-Kinvariance-WGamma}.
\end{enumerate}
The first two types of invariance are free of the cutoff function $R$.
In fact, the invariance of the Wilson action under translation and
rotation can also be written without $R$ \cite{Sonoda:2015pva}:
\begin{subequations}
\begin{eqnarray}
\int_p D_\mu^T (p) \phi (p) \cdot \frac{\delta S[\phi]}{\delta \phi
  (p)} &=& 0\,,\\
\int_p D^R_{\mu\nu} (p) \phi (p) \cdot \frac{\delta S[\phi]}{\delta \phi
  (p)} &=& 0\,.
\end{eqnarray}
\end{subequations}
On the other hand, the invariance under the scale and special
conformal transformations depends non-trivially on the cutoff function
$R$.

As for the special conformal invariance, Eq.~(\ref{Kinvariance-Gamma})
for $\Gamma$ has been obtained by Rosten as (4.16) in
\cite{Rosten:2016zap}.  A similar expression has also been derived as
(10) in \cite{Delamotte:2015aaa}.

\section{Wilson-Fisher fixed point to order $\ep$\label{section-example}}

As a concrete example, we consider the Wilson-Fisher fixed point in $D
= 4 - \ep$ dimensions, and construct a conformally invariant 1PI
action $\Gamma$ to first order in $\ep$.  Assuming $\gamma = 0$ at
this order, we obtain the following equations from
(\ref{Sinvariance-WGamma}) and (\ref{Kinvariance-WGamma}):
\begin{enumerate}
\item Scale invariance 
\begin{equation}
\int_p \left( \frac{D+2}{2} + p \cdot \partial_p \right) \Phi (p)
\cdot \frac{\delta \Gamma [\Phi]}{\delta \Phi (p)} + \int_p \left(2 - p
  \cdot \partial_p \right) R(p) \cdot \frac{1}{2} G_{p,-p} [\Phi] =
0\,.
\label{Gamma-S}
\end{equation}
\item Special conformal invariance
\begin{eqnarray}
&&\int_p \left( p_\nu \frac{\partial^2}{\partial p_\mu \partial p_\nu} -
  \frac{1}{2} p_\mu \frac{\partial^2}{\partial p_\nu \partial p_\nu} +
  \frac{D+2}{2} \frac{\partial}{\partial p_\mu} \right) \Phi (p) \cdot
\frac{\delta \Gamma [\Phi]}{\delta \Phi (p)}\nn\\
&&\quad + \int_p \left(2 - p
  \cdot \partial_p \right) R(p) \cdot \frac{1}{2}
\frac{\partial}{\partial p_\mu} G_{-p,q} [\Phi]\Big|_{q=p} = 0\,.
\label{Gamma-K}
\end{eqnarray}
\end{enumerate}
We will solve these equations with the ansatz
\begin{equation}
\Gamma [\Phi] = - \frac{1}{2} \int_p \left( p^2 + m^2 \right) \Phi (p)
\Phi (-p) - \lambda \frac{1}{4!} \int_{p_1,\cdots,p_4} \Phi (p_1)
\cdots \Phi (p_4)\, \delta (p_1 + \cdots + p_4)\,,\label{Gamma-ansatz}
\end{equation}
where $m^2, \lambda$ are both of order $\ep$.  Note this is automatically
invariant under translation and rotation.

The high momentum propagator $G_{-p,q} [\Phi]$ is now defined by
\begin{eqnarray}
&&\int_q G_{-p, q} [\Phi] \left( \left(q^2 + m^2 + R (q) \right) \delta
  (q-r) + \frac{\lambda}{2} \int_{p_1, p_2} \Phi (p_1) \Phi (p_2)
  \delta (p_1+p_2-q+r) \right) \nn\\
&&\qquad = \delta (p-r)\,,
\end{eqnarray}
and it is obtained as
\begin{eqnarray}
G_{-p,q} [\Phi] &=& \frac{1}{p^2 + R(p)} \delta (p-q)
 - \frac{m^2}{(p^2 + R(p))^2} \delta (p-q) \nn\\
&&- \lambda \frac{1}{p^2 + R(p)} \frac{1}{q^2 + R(q)}
\frac{1}{2} \int_{p_1, p_2} \Phi (p_1) \Phi (p_2) \delta
(p_1+p_2 - p + q)
\end{eqnarray}
up to first order in $\ep$.

\subsection{Scale invariance}

Substituting (\ref{Gamma-ansatz}) into (\ref{Gamma-S}), we obtain two
equations, one quadratic in $\Phi$, and the other quartic in $\Phi$.
The latter is given by
\begin{equation}
\frac{\lambda}{4!} \int_{p_1,\cdots,p_4} \Phi (p_1) \cdots \Phi (p_4)
\sum_{i=1}^4 \left( \frac{D-2}{2} + p_i \cdot \partial_{p_i} \right)
\delta (p_1+\cdots+p_4) = 0\,.
\end{equation}
This gives
\begin{equation}
\lambda \, (4-D) = 0\,, \label{quartic}
\end{equation}
which is trivially satisfied to order $\ep$.  We are now left with
\begin{equation}
\frac{1}{2} \int_p \Phi (p) \Phi (-p) \lb
(2 - p \cdot \partial_p) \left(p^2 + m^2 \right) + \frac{\lambda}{2}
\int_q \frac{ (2 - q \cdot \partial_q) R(q)}{(q^2 + R(q))^2} \rb = 0\,.
\end{equation}
This is solved by
\begin{equation}
m^2 = - \frac{\lambda}{4} \int_q \frac{ (2 - q \cdot \partial_q)
  R(q)}{(q^2 + R(q))^2} \,.\label{quadratic}
\end{equation}

\subsection{Special conformal invariance}

Substituting (\ref{Gamma-ansatz}) into (\ref{Gamma-K}), we obtain
two equations, one quadratic in $\Phi$, and the other quartic in $\Phi$.
The latter is given by
\begin{eqnarray}
&&\frac{\lambda}{4!} \int_{p_1,\cdots,p_4} \delta (p_1+\cdots+p_4)\nn\\
&&\quad \times \sum_{i=1}^4 \left( p_{i\nu} \frac{\partial^2}{\partial
    p_{i\mu} \partial p_{i\nu}} - \frac{1}{2} p_{i\mu}
  \frac{\partial^2}{\partial p_{i\nu} \partial p_{i\nu}} +
  \frac{D+2}{2} \frac{\partial}{\partial p_{i\mu}} \right) \cdot \Phi (p_1) 
\cdots \Phi (p_4) = 0\,.
\end{eqnarray}
Using
\begin{equation}
\frac{\partial}{\partial p_{i\mu}} \delta (p_1+\cdots + p_4) =
\frac{\partial}{\partial p_{1\mu}} \delta (p_1+\cdots + p_4)
\end{equation}
(independent of $i$), we obtain
\begin{equation}
\frac{\lambda}{4!} \int_{p_1,\cdots,p_4}  \Phi (p_1) 
\cdots \Phi (p_4) (D-4) \frac{\partial}{\partial p_{1\mu}} \delta
(p_1+\cdots +p_4) = 0\,,
\end{equation}
which gives (\ref{quartic}) again.  The equation quadratic in $\Phi$
is given by
\begin{eqnarray}
&&\int_p \Phi (-p) (p^2+m^2) \left( p_\nu \frac{\partial^2}{\partial
    p_\mu \partial p_\nu} - \frac{1}{2} p_\mu
  \frac{\partial^2}{\partial p_\nu \partial p_\nu} + \frac{D+2}{2}
  \frac{\partial}{\partial p_\mu} \right) \Phi (p)\nn\\
&&- \lambda \frac{1}{2} \int_p \left(- p \cdot \partial_p + 2\right)
R(p) \cdot \frac{1}{(p^2+R(p))^2} \frac{1}{2} \int_{p_1, p_2} \Phi
(p_1) \Phi (p_2)
\frac{\partial}{\partial p_{1\mu}} \delta (p_1+p_2) = 0\,.
\end{eqnarray}
Integration by parts reduces this to
\begin{equation}
  \int_p \frac{\partial}{\partial p_\mu} \Phi (p) \cdot \Phi (-p) \left(
    2 m^2 + \lambda \frac{1}{2} \int_q \frac{\left(2 -q
        \cdot \partial_q \right) R(q)}{(q^2+R(q))^2} \right) = 0 \,.
\end{equation}
Hence, we obtain (\ref{quadratic}) again.  We have thus seen that
scale invariance automatically leads to conformal invariance. 

We need a second order calculation to fix $\lambda$ to order $\ep$.
(It turns out $\frac{\lambda}{(4\pi)^2} =
\frac{\ep}{3}$. \cite{Wilson:1973jj})

\section{Conformal transformation of composite operators\label{section-composite}}

Let $\Op (p)$ be a scalar composite operator of scale dimension $- y$
with momentum $p$.  (In coordinate space the scale dimension is
$-y+D$.)  Translations and rotations act on $\Op (p)$ the same way as
on $\phi (p)$; we only need to generalize $D^S (p)$ and $D^K_\mu (p)$
as
\begin{subequations}
\begin{eqnarray}
D^S (p) \Op (p) &\equiv& \left( - p_\mu \frac{\partial}{\partial p_\mu} - y
\right) \Op (p)\,,\\
D^K_\mu (p) \Op (p) &\equiv& \left( - p_\nu \frac{\partial^2}{\partial
    p_\mu \partial p_\nu} + \frac{1}{2} p_\mu
  \frac{\partial^2}{\partial p_\nu \partial p_\nu} - y
  \frac{\partial}{\partial p_\mu} \right) \Op (p)\,.
\end{eqnarray}
\end{subequations}
The invariance under scale and special conformal transformations is
now given by
\begin{subequations}
\label{comp-invariance}
\begin{eqnarray}
D^S (p) \Op (p) - e^{-S} \int_q K(q) \frac{\delta}{\delta \phi (q)} \left(
D^S (q) \left[ \Op (p) \Phi (q)\right] \,e^S \right)&=& 0\,,\\
D^K_\mu (p) \Op (p) - e^{-S} \int_q K(q) \frac{\delta}{\delta \phi (q)} \left(
D^K_\mu (q) \left[ \Op (p) \Phi (q)\right] \,e^S \right)&=& 0\,,
\end{eqnarray}
\end{subequations}
where the product of composite operators is defined by
\begin{equation}
\left[ \Op (p) \Phi (q) \right] \equiv  \Op (p) \Phi (q) +
  \frac{k(q)}{q^2 K(q)} 
  \frac{\delta \Op (p)}{\delta \phi (-q)}\,.
\end{equation}
Eqs.~(\ref{comp-invariance}) imply
\begin{subequations}
\begin{eqnarray}
&& D^S (p) \vvev{\Op (p)\, \phi (p_1)
  \cdots \phi (p_n)}
+ \sum_{i=1}^n D^S (p_i) \,\vvev{\Op (p)\,\phi (p_1) \cdots \phi
  (p_n)} = 0\,,\\
&& D^K_\mu (p) \vvev{\Op (p)\, \phi (p_1)
  \cdots \phi (p_n)}
+ \sum_{i=1}^n D^K_\mu (p_i) \,\vvev{\Op (p)\,\phi (p_1) \cdots \phi
  (p_n)} = 0\,.
\end{eqnarray}
\end{subequations}

For completeness let us rewrite (\ref{comp-invariance}) in terms of
$W[J]$ and $\Gamma [\Phi]$.  Regarding $\Op (p)$ as a functional of
$J$, we obtain
\begin{subequations}
\begin{eqnarray}
&&\left(- p \cdot \partial_p - y\right) \Op (p) + \int_q J(q) \left( - q
     \cdot \partial_q - \frac{D+2}{2} + \gamma \right) \frac{\delta
     \Op (p)}{\delta J(q)}\nn\\
&&\quad + \int_q \left(- q \cdot \partial_q + 2 - 2 \gamma\right) R (q)
     \cdot
\lb \frac{\delta W[J]}{\delta J(-q)} \frac{\delta \Op (p)}{\delta
     J(q)} + \frac{1}{2} \frac{\delta^2 \Op (p)}{\delta J(-q) \delta
     J(q)} \rb = 0\,,\\
&&D_\mu^K (p) \Op (p) + \int_q J(-q) D_\mu^K (q) \frac{\delta \Op
     (p)}{\delta J(-q)}
 + \frac{1}{2} \int_q \left( - q \cdot \partial_q + 2 - 2 \gamma
     \right) R (q) \nn\\
&&\quad \cdot \frac{\partial}{\partial q_\mu} \left(
\frac{\delta^2 \Op (p)}{\delta J(-q) \delta J(q')} + \frac{\delta W}{\delta
     J(-q)} \frac{\delta \Op (p)}{\delta J(q')} + \frac{\delta
     W}{\delta J(q')} \frac{\delta \Op (q)}{\delta
     J(-q)}\right)\Big|_{q'=q} = 0\,.
\end{eqnarray}
\end{subequations}
Alternatively, regarding $\Op (p)$ as a functional of $\Phi$, we obtain
\begin{subequations}
\begin{eqnarray}
&&\left( - p \cdot \partial_p - y \right) \Op (p) - \int_q \left(- q
     \cdot \partial_q - \frac{D+2}{2} + \gamma\right) \Phi (q) \cdot
     \frac{\delta \Op (p)}{\delta \Phi (q)}\nn\\
&&\quad + \int_q \left( - q \cdot \partial_q + 2 - 2 \gamma\right) R
     (q)\cdot 
\frac{1}{2} \int_{r,s} G_{q,-r} \frac{\delta^2 \Op (p)}{\delta \Phi
     (r) \delta \Phi (-s)} G_{s,-q} = 0\,,\\
&&D_\mu^K (p) \Op (p) - \int_q D_\mu^K (q) \Phi (q) \cdot \frac{\delta
     \Op (p)}{\delta \Phi (q)}\nn\\
&&\quad + \frac{1}{2} \int_q \left( - q \cdot \partial_q + 2 - 2 \gamma\right) R
     (q) \cdot \int_{r,s} \frac{\partial G_{-q,r}}{\partial q_\mu}
     \frac{\delta^2 \Op (p)}{\delta \Phi (-r) \delta \Phi (s)}  G_{-s,q}= 0\,.
\end{eqnarray}
\end{subequations}
It is the easiest to obtain the above results by varying either $W$ or
$\Gamma$ infinitesimally by $\Op (p)$ in (\ref{Sinvariance-WGamma})
and (\ref{Kinvariance-WGamma}).

A concrete example is
\begin{equation}
\left[\frac{1}{2} \phi^2 (p)\right] \equiv \frac{1}{2} \int_{p_1, p_2}
\Phi (p_1) \Phi (p_2) \delta (p_1+p_2-p) + \kappa_2 \delta (p)
\end{equation}
at the Gaussian fixed point in $D > 2$.  With $y = 2$, both of
(\ref{comp-invariance}) are satisfied if the constant
$\kappa_2$ is chosen as
\begin{equation}
\kappa_2 = - \frac{1}{2(D-2)} \int_p
\frac{(2-p\cdot \partial_p) R(p)}{(p^2+R(p))^2}\,.
\end{equation}

\section{Conclusion\label{conclusion}}

The main purpose of this paper is to reformulate the recent results of
Rosten \cite{Rosten:2014oja, Rosten:2016nmc, Rosten:2016zap} using the
method of equation-of-motion composite operators advocated in
\cite{Igarashi:2009tj}.  The Wilson action of the continuum limit of a
theory has all the symmetry intact despite the presence of a finite
momentum cutoff.  We hope that we have convinced the reader that a
finite UV cutoff does not stand in the way of making a Wilson action
invariant under conformal transformations.

Note added: Rosten extends his work further in a recent article
\cite{Rosten:2017}.

\appendix

\section{Quick summary of the ERG formalism\label{appendix-ERG}}

The purpose of this and next appendices is to give the reader (without
the working knowledge of ERG) just enough to follow the flow of the
present paper.  For further details we recommend
\cite{Igarashi:2016qdr} and references cited therein.

As in the main text, we use the dimensionless notation in which
dimensionful quantities are measured in units of an appropriate power
of the momentum cutoff.  Hence, the momentum cutoff becomes $1$ in
this convention.

The renormalization group flow of the Wilson action $S_t [\phi]$ is
given by the exact renormalization group equation\cite{Wilson:1973jj}
\begin{eqnarray}
  \partial_t e^{S_t} &=&\int_p \left(- p_\mu \frac{\partial}{\partial
      p_\mu} \ln K(p) + 
    \frac{D+2}{2} - \gamma_t + p_\mu \frac{\partial}{\partial p_\mu}
  \right) \phi (p) \cdot \frac{\delta}{\delta \phi (p)}\, e^{S_t}\nn\\
  &&+ \int_p \left( - p_\mu \frac{\partial}{\partial p_\mu} \ln
    \frac{K(p)^2}{k(p)} - 2 \gamma_t \right) \frac{k(p)}{p^2} \frac{1}{2}
    \frac{\delta^2}{\delta \phi (-p) \delta \phi (p)}\, e^{S_t}\,,
\end{eqnarray}
where $t$ is the logarithmic scale factor.  This is a generalized
version with two cutoff functions $K(p),\,k(p)$\cite{Sonoda:2015bla}:
$K(p)$ approaches $1$ as $p \to 0$, and decreases rapidly for $p \gg
1$, and $k(p)$ vanishes at $p = 0$.  In the popular adaptation by
Polchinski\cite{Polchinski:1983gv}, $k(p)$ is taken as
\begin{equation}
k(p) = K(p) \left(1 - K(p)\right)\,.
\end{equation}
To obtain $S_{t+\Delta t}$ from $S_t$, we first integrate over the
field with momenta between $1$ and $e^{-\Delta t}$.  We then rescale
the momentum by the factor $e^{\Delta t}$ to restore the cutoff at $1$,
and renormalize the field so that, for example, the kinetic term is
canonically normalized.  It is remarkable that this whole procedure can be
expressed as a functional differential equation.

In this paper we are not interested in $t$-dependent actions, but only
interested in a fixed point solution $S [\phi]$, satisfying
\begin{eqnarray}
  0 &=&\int_p \left(- p_\mu \frac{\partial}{\partial
      p_\mu} \ln K(p) + 
    \frac{D+2}{2} - \gamma + p_\mu \frac{\partial}{\partial p_\mu}
  \right) \phi (p) \cdot \frac{\delta}{\delta \phi (p)}\, e^{S}\nn\\
  &&+ \int_p \left (- p_\mu \frac{\partial}{\partial p_\mu} \ln
    \frac{K(p)^2}{k(p)} - 2 \gamma \right) \frac{k(p)}{p^2} \frac{1}{2}
    \frac{\delta^2}{\delta \phi (-p) \delta \phi (p)}\, e^{S}\,,
\end{eqnarray}
where $\gamma$ is a constant anomalous dimension.  This $S$ has a UV
cutoff $p = 1$, just like a generic bare action with the same cutoff
$p = 1$, but it corresponds to a massless continuum theory.  The field
with momenta $p > 1$ have already been integrated, and the Wilson
action can provide the continuum limit of correlation functions 
only with a little modification\cite{Sonoda:2015bla}:
\begin{eqnarray}
\vvev{\phi (p_1) \cdots \phi (p_n)} &\equiv&
\prod_{i=1}^n \frac{1}{K(p_i)} \cdot \vev{\exp \left( - \frac{1}{2}
    \int_p \frac{k(p)}{p^2} \frac{\delta^2}{\delta \phi (p)\delta \phi
      (-p)} \right) \phi (p_1) \cdots \phi (p_n)}\\
&=& \prod_{i=1}^n \frac{1}{K(p_i)} \cdot \int [d\phi]\, e^{S}
\exp \left( - \frac{1}{2}
    \int_p \frac{k(p)}{p^2} \frac{\delta^2}{\delta \phi (p)\delta \phi
      (-p)} \right) \phi (p_1) \cdots \phi (p_n)\,.\nn
\end{eqnarray}
$k(p)$ modifies the two-point functions trivially at high momenta, and
$K(p)$ corrects the normalization of the field.  As befits the
continuum limit, the modified correlation functions are defined for
arbitrary momenta, and satisfy the scaling law
\begin{equation}
\vvev{\phi (p_1 e^t) \cdots \phi (p_n e^t)} = \exp \left( n \left(-
    \frac{D+2}{2} + \gamma \right) t \right) \vvev{\phi (p_1) \cdots
  \phi (p_n)}\,.
\end{equation}
Hence, the two-point function is given by
\begin{equation}
\vvev{\phi (p) \phi (q)} = \frac{\mathrm{const}}{p^{2(1-\gamma)}}
\,\delta (p+q)\,.
\end{equation}

We next introduce the concept of composite operators.  (For more
details than given here, see Sect.~4 of \cite{Igarashi:2009tj}.)  A composite
operator $\Op (p)$ is a functional of $\phi$, and it can be regarded
as an infinitesimal variation of the action.  We define its modified
correlation functions by
\begin{equation}
\vvev{\Op (p) \, \phi (p_1) \cdots \phi (p_n)}
\equiv \prod_{i=1}^n \frac{1}{K(p_i)} \cdot \vev{\Op (p)\exp \left( -
    \frac{1}{2} \int_q \frac{k(q)}{q^2} \frac{\delta^2}{\delta \phi
      (-q) \delta \phi (q)}\right) \phi (p_1) \cdots \phi (p_n)}\,.
\label{appendix-compositeoperator}
\end{equation}
Note the absence of $K(p)$ for the composite operator.  There are two
special composite operators playing important roles in this paper.
One is
\begin{equation}
\Phi (p) \equiv \frac{1}{K(p)} \left( \phi (p) + \frac{k(p)}{p^2}
  \frac{\delta S}{\delta \phi (-p)} \right) \label{appendix-Phi}
\end{equation}
which has the correlation functions
\begin{equation}
\vvev{\Phi (p) \phi (p_1) \cdots \phi (p_n)} = \vvev{\phi (p) \phi
  (p_1) \cdots \phi (p_n)}\,.
\end{equation}
$\Phi (p)$ is a composite operator, but it shares the same modified
correlation functions as the elementary field $\phi (p)$.  The other
is a special class of composite operators, called equation-of-motion
composite operators (a.k.a. redundant operators).  They are given in
the form
\begin{equation}
\E_\Op \equiv - e^{-S} \int_p K(p) \frac{\delta}{\delta \phi (p)}
\left( \Op (p) e^S \right)\,,
\end{equation}
where $\Op (p)$ is a composite operator.  $\E_\Op$ has the correlation
functions
\begin{equation}
\vvev{\E_\Op\, \phi (p_1) \cdots \phi (p_n)}
= \sum_{i=1}^n \vvev{\phi (p_1) \cdots \Op (p_i) \cdots \phi (p_n)}\,.
\end{equation}
(Derivation) Using (\ref{appendix-compositeoperator}), we obtain
\begin{eqnarray}
&&\vvev{\E_\Op\, \phi (p_1) \cdots \phi (p_n)}
\equiv \prod_{i=1}^n \frac{1}{K(p_i)} \int_p K(p)\nn\\
&&\times \vev{e^{-S} (-) \frac{\delta}{\delta \phi (p)} \left( \Op
    (p) e^S \right)\, \exp \left(-\frac{1}{2} \int_q
    \frac{k (q)}{q^2}\frac{\delta^2}{\delta \phi (q)\delta \phi
      (-q)}\right) \phi (p_1) \cdots \phi (p_n)} \,.
\end{eqnarray}
Functionally integrating this by part, we obtain
\begin{eqnarray}
&&\vvev{\E_\Op\, \phi (p_1) \cdots \phi (p_n)}
= \prod_{i=1}^n \frac{1}{K(p_i)} \int_p K(p) \nn\\
&&\quad \times \vev{ \Op (p) \exp \left(-\frac{1}{2} \int_q
    \frac{k (q)}{q^2}\frac{\delta^2}{\delta \phi (q)\delta \phi (-q)} \right)
\frac{\delta}{\delta \phi (p)}\lb \phi (p_1) \cdots \phi (p_n)\rb } \nn\\
&&= \prod_{i=1}^n \frac{1}{K(p_i)} \sum_{j=1}^n K(p_j)
\vev{\Op (p_j) \exp \left(-\frac{1}{2} \int_q
    \frac{k (q)}{q^2}\frac{\delta^2}{\delta \phi (q)\delta \phi (-q)} \right)
\phi (p_1) \cdots \widehat{\phi (p_j)} \cdots \phi (p_n)}\nn\\
&&= \sum_{i=1}^n \vvev{\Op (p_i) \phi (p_1)
  \cdots \widehat{\phi (p_i)} \cdots \phi (p_n)}\,,
\end{eqnarray}
where the hat above $\phi$ implies the omission. (End of derivation)

Given two composite operators $\Op_1 (p), \Op_2 (q)$, their product
$\Op_1 (p) \Op_2 (q)$ is not necessarily a composite operator.  When
one of them is $\Phi (p)$, however, its product with an arbitrary $\Op
(q)$ is easy to construct:
\begin{eqnarray}
\left[ \Phi (p) \Op (q)  \right] &\equiv&
 \Phi (p) \Op (q) + \frac{k(p)}{p^2 K(p)} \frac{\delta \Op (q)}{\delta \phi
  (-p)} \nn\\
&=& e^{-S} \frac{1}{K(p)} \left( \phi (p) + \frac{k(p)}{p^2}
  \frac{\delta}{\delta \phi (-p)} \right) \left( \Op (q) e^S \right)\,.
\end{eqnarray}
The product has the correlation functions
\begin{equation}
\vvev{\left[\Phi (p) \Op (q) \right]\, \phi (p_1) \cdots \phi (p_n)}
= \vvev{\Op (q)\,\phi (p) \phi (p_1) \cdots \phi (p_n)}\,.
\end{equation}

\section{Generating functional $W[J]$ and 1PI action $\Gamma [\Phi]$\label{appendix-WGamma}}

We can interpret a Wilson action $S[\phi]$ as a generating functional
of the connected correlation functions of the scalar field for which
only the field with momentum higher than the cutoff $p=1$ has been
integrated.  Regarding
\begin{equation}
J(p) \equiv \frac{R(p)}{K(p)} \phi (p) \label{appendix-J}
\end{equation}
as the source, we obtain the generating functional as
\begin{equation}
W[J] \equiv S [\phi] + \frac{1}{2} \int_p \frac{1}{R(p)} J(p)
J(-p)\,,\label{appendix-W} 
\end{equation}
where
\begin{equation}
R (p) \equiv \frac{p^2}{k(p)} K(p)^2\,.
\end{equation}
Recall that $S[\phi]$ depends on two cutoff functions $K$ \& $k$, but
$W [J]$ and $\Gamma [\Phi]$, to be defined shortly, depend only on
this $R$.

It is straightforward to check that the composite operator $\Phi (p)$,
defined by (\ref{appendix-Phi}), is obtained as
\begin{equation}
\Phi (p) = \frac{\delta W[J]}{\delta J(-p)}\,.\label{appendix-Phi2}
\end{equation}
The 1PI action $\Gamma [\Phi]$ is now defined as the Legendre
transform of the generating functional $W[J]$ as
\begin{equation}
\Gamma [\Phi] - \frac{1}{2} \int_p R(p) \Phi (p) \Phi (-p) = W[J] -
\int_p J(p) \Phi (-p)\,.\label{appendix-Gamma}
\end{equation}
Differentiating this with respect to $\Phi (-p)$, we obtain
\begin{equation}
J(p) = R(p) \Phi (p) - \frac{\delta \Gamma [\Phi]}{\delta \Phi
  (-p)}\,.\label{appendix-J2}
\end{equation}
The high momentum propagator, defined by
\begin{equation}
G_{p, q} [\Phi] \equiv \frac{\delta^2 W[J]}{\delta J(p) \delta
  J(q)}\,,\label{appendix-G} 
\end{equation}
is symmetric with respect to $p$ \& $q$, and satisfies
\begin{equation}
\int_q G_{p,q} [\Phi] \left( R (q) \delta (q-r) - \frac{\delta^2
    \Gamma [\Phi]}{\delta \Phi (-q) \delta \Phi (-r)} \right) = \delta
(p-r)\,.
\end{equation}

Consider the simplest example of the Gaussian fixed point:
\begin{equation}
S_G [\phi] \equiv - \frac{1}{2} \int_p \frac{p^2}{K(p)^2+k(p)} \phi (p)
\phi (-p)\,.
\end{equation}
We obtain
\begin{eqnarray}
W_G [J] &=& \frac{1}{2} \int_p \frac{1}{p^2 + R(p)} J(p) J(-p)\,,\\
\Gamma_G [\Phi] &=& - \frac{1}{2} \int_p p^2 \Phi (p) \Phi (-p)\,.
\end{eqnarray}
Hence, the high momentum propagator is given by
\begin{equation}
G_{p,q} [\Phi] = \frac{1}{p^2 + R(p)}\, \delta (p+q)\,.
\end{equation}
It is trivial to check
\begin{subequations}
\begin{eqnarray}
\vev{\phi (p) \phi (q)}_G &=& \frac{K(p)^2 + k(p)}{p^2} \delta (p+q)\,,\\
\vvev{\phi (p) \phi (q)}_G &=& \frac{1}{K(p) K(q)} \left( \vev{\phi (p)
                           \phi (q)} - \frac{k(p)}{p^2} \delta (p+q)
                           \right)\nn\\
&=& \frac{1}{p^2}\, \delta (p+q)\,.
\end{eqnarray}
\end{subequations}

\section{Derivation of (\ref{conformal-invariance}) from the
  energy-momentum tensor\label{appendix-EM}}

As has been shown in \cite{Polchinski:1987dy}, conformal invariance is
equivalent to the vanishing of the trace of the energy-momentum
tensor; scale invariance equivalent to the vanishing of its integral.
It is therefore natural that the author of \cite{Rosten:2014oja} was
led to consider the energy-momentum tensor in the realization of
conformal algebra for Wilson actions.  In this appendix we wish to
summarize how to derive (\ref{conformal-invariance}) from the relevant
properties of the energy-momentum tensor.  We will follow
\cite{Sonoda:2015pva}, since we can obtain the particular form of
$\Sigma$'s given by (\ref{Sigma-definition}) without any effort.

Now, in \cite{Sonoda:2015pva} we have assumed the invariance of the Wilson
action under translations and rotations
\begin{subequations}
\label{appendix-TR}
\begin{eqnarray}
\Sigma^T_\mu &\equiv& - e^{-S} \int_p K(p) \frac{\delta}{\delta \phi
  (p)} \left( D^T_\mu (p) \phi (p) \, e^S\right) =
0\,,\label{appendix-T}\\ 
\Sigma^R_{\mu\nu} &\equiv& - e^{-S} \int_p K(p) \frac{\delta}{\delta
  \phi (p)} \left( D^R_{\mu\nu} (p) \phi (p)\, e^S \right) =
0\,,\label{appendix-R} 
\end{eqnarray}
\end{subequations}
where $D^T_\mu, D^R_{\mu\nu}$ are defined in (\ref{D-definition}).  We
have then shown the existence of the energy-momentum tensor
$\Theta_{\mu\nu} (p)$ satisfying
\begin{subequations}
\label{EM-TR}
\begin{eqnarray}
p_\mu \Theta_{\mu\nu} (p) &=& \int_q K(q) e^{-S} \frac{\delta}{\delta
  \phi (q)} \left( (p+q)_\nu \Phi (p+q) e^S \right)\,,\label{EM-T}\\
\Theta_{\mu\nu} (p) &=& \Theta_{\nu\mu} (p)\,.\label{EM-R}
\end{eqnarray}
\end{subequations}
It is straightforward to go backward, and derive (\ref{appendix-TR})
from (\ref{EM-TR}).  To obtain (\ref{appendix-T}), we simply set $p=0$
in (\ref{EM-T}).  To obtain (\ref{appendix-R}), differentiate
(\ref{EM-T}) with respect to $p_\alpha$, antisymmetrize the result
with respect to $\nu$ \& $\alpha$, and set $p=0$.

The invariance under scale and special conformal transformations is
given respectively by
\begin{subequations}
\label{appendix-SK}
\begin{eqnarray}
\Sigma^S &\equiv& - e^{-S} \int_p K(p) \frac{\delta}{\delta \phi (p)}
\left( D^S (p) \Phi (p) \,e^S \right) = 0\,,\label{appendix-S}\\
\Sigma^K_\mu &\equiv& - e^{-S} \int_p K(p) \frac{\delta}{\delta \phi
  (p)} \left( D^K_\mu (p) \Phi (p)\, e^S \right) = 0\,,\label{appendix-K}
\end{eqnarray}
\end{subequations}
where $D^S, D^K_\mu$ are defined in (\ref{D-definition}).
We wish to show how to obtain these from the trace condition:
\begin{equation}
\Theta (p) \equiv \Theta_{\mu\mu} (p) =  \left(\frac{D-2}{2} + \gamma
\right) \int_q K(q) e^{-S} 
\frac{\delta}{\delta \phi (q)} \left( \Phi (p+q) e^S \right)\,. \label{EM-SK}
\end{equation}
In \cite{Sonoda:2015pva} it is shown that a fixed point Wilson
action, satisfying (\ref{appendix-S}), also satisfies (\ref{EM-SK}) at
$p=0$.   Conversely, to obtain (\ref{appendix-S}) from (\ref{EM-TR}) and
(\ref{EM-SK}),
we differentiate (\ref{EM-T}) with respect to $p_\nu$, sum over
$\nu$, and then set $p=0$ to obtain
\begin{equation}
\Theta (0) = \int_q K(q) e^{-S} \frac{\delta}{\delta \phi (q)} \left(
  (D + q \cdot \partial_q) \Phi (q) \, e^S \right)\,.
\end{equation}
Using (\ref{EM-SK}) with $p=0$, we obtain(\ref{appendix-S}). 

Getting (\ref{appendix-K}) from (\ref{EM-TR}) \& (\ref{EM-SK}) is a
little more involved.  (This has been done in Sect.~VI of
\cite{Sonoda:2015pva}, where (\ref{EM-SK}) is assumed up to a
two-derivative term $p_\mu p_\nu L_{\mu\nu} (p)$.  For simplicity, we
have removed the two-derivative term by redefining $\Theta_{\mu\nu}
(p)$.)  We apply
\[
\frac{\partial^2}{\partial p_\alpha \partial p_\nu} - \frac{1}{2}
\delta_{\alpha\nu} \frac{\partial^2}{\partial p_\beta \partial
  p_\beta}
\]
on (\ref{EM-T}) and set $p=0$.  Using (\ref{EM-SK}), we can write the
left side as
\begin{equation}
\frac{\partial}{\partial p_\alpha} \Theta (p)\Big|_{p=0} =
\left(\frac{D-2}{2}+\gamma \right) \int_q K(q) e^{-S} 
\frac{\delta}{\delta \phi (q)} \left( \frac{\partial}{\partial
    q_\alpha} \Phi (q)\,e^S \right)\,.
\end{equation}
The right side gives
\begin{equation}
\int_q K(q) e^{-S} \frac{\delta}{\delta \phi (q)} \lb
\left( q_\nu \frac{\partial^2}{\partial q_\alpha \partial q_\nu} -
  \frac{1}{2} q_\alpha \frac{\partial^2}{\partial q_\beta \partial
    q_\beta} + D \frac{\partial}{\partial q_\alpha} \right) \Phi (q)
\, e^S \rb\,.
\end{equation}
Equating the two sides, we obtain $\Sigma_\alpha^K = 0$.

In a recent work \cite{Rosten:2016nmc} Rosten regards (\ref{EM-TR}) and
(\ref{EM-SK}) as fundamental equations from which he attempts to
construct a conformally invariant Wilson action.

\section{Derivation of (\ref{Kinvariance-WGamma})\label{appendix-Kinvariance-WGamma}}

We wish to rewrite the special conformal invariance $\Sigma^K_\mu =
0$, where $\Sigma^K_\mu$ is defined by (\ref{SigmaK-definition}), in
terms of the generating functional $W[J]$ and 1PI action $\Gamma
[\Phi]$.  (The content of this appendix overlaps with the main subject
of \cite{Rosten:2016zap}.  Our Wilson action is more simply related to
$\Gamma$, resulting in a simpler derivation.)  We first expand
$\Sigma^K_\mu$ as
\begin{equation}
\Sigma_\mu^K = - \int_p K(p) D_\mu^K (p) \left(\frac{\delta \Phi
    (p)}{\delta \phi (q)} + \Phi (p) \frac{\delta S [\phi]}{\delta \phi (q)}
\right)\Big|_{q=p}\,,
\end{equation}
where we set $q=p$ only after the action of $D_\mu^K (p)$.
Then, using
\begin{equation}
\Phi (p) = \frac{\delta W[J]}{\delta J(-p)},\quad
J(p) = \frac{R(p)}{K(p)}\,\phi (p)\,,\quad
S[\phi] = W[J] - \frac{1}{2} \int_p \frac{J(p) J(-p)}{R(p)}\,,
\end{equation}
we obtain
\begin{eqnarray}
\Sigma_\mu^K &=& - \int_p R(p) \lb D_\mu^K (p) \frac{\delta^2 W[J]}{\delta
  J(q)\delta J(-p)} \Big|_{q=p} + D_\mu^K (p) \frac{\delta
  W[J]}{\delta J(-p)} \cdot \left(\frac{\delta W[J]}{\delta J(p)} -
  \frac{J(-p)}{R(p)}\right) \rb\nn\\
&=& \int_p J(-p) D_\mu^K (p) \frac{\delta W[J]}{\delta J(-p)} \nn\\
&& \quad - \int_p R(p) D_\mu^K (p) \lb \frac{\delta^2 W[J]}{\delta
  J(q)\delta J(-p)} + \frac{\delta W[J]}{\delta J(q)} \frac{\delta
  W[J]}{\delta J(-p)} \rb_{q=p}\,.
\end{eqnarray}
To transform the last integral, we use a formula of partial integration
\begin{eqnarray}
&&\int_p R(p) \lb p_\nu \frac{\partial^2}{\partial p_\mu \partial p_\nu}
- \frac{1}{2} p_\mu \frac{\partial^2}{\partial p_\nu \partial p_\nu}
\rb F(-p,q)\Big|_{q=p}\nn\\
&&= - \frac{1}{2} \int_p \left(D + p \cdot \partial_p\right) R (p) \cdot
\frac{\partial}{\partial p_\mu} F(-p,q)\Big|_{q=p}\label{formula}
\end{eqnarray}
which is valid for any symmetric $F (-p,q)$ satisfying
\begin{equation}
F(-p,q) = F (q, -p)\,.
\end{equation}
We then obtain
\begin{eqnarray}
&&- \int_p R(p)  D_\mu^K (p) \lb \frac{\delta^2 W[J]}{\delta
  J(q)\delta J(-p)} + \frac{\delta W[J]}{\delta J(q)} \frac{\delta
  W[J]}{\delta J(-p)} \rb_{q=p}\nn\\
&&= \frac{1}{2} \int_p ( - p \cdot \partial_p + 2 - 2 \gamma) R(p) \cdot
\frac{\partial}{\partial p_\mu}  \lb \frac{\delta^2 W[J]}{\delta
  J(q)\delta J(-p)} + \frac{\delta W[J]}{\delta J(q)} \frac{\delta
  W[J]}{\delta J(-p)} \rb_{q=p}\,.
\end{eqnarray}
Hence, (\ref{Kinvariance-W}) is obtained:
\begin{eqnarray}
&&\int_p J(-p) D_\mu^K (p) \frac{\delta W[J]}{\delta J(-p)}\nn\\
&&\quad + \frac{1}{2} \int_p  (- p \cdot \partial_p +2 - 2 \gamma) R(p) \cdot
\frac{\partial}{\partial p_\mu}  \lb \frac{\delta^2 W[J]}{\delta
  J(q)\delta J(-p)} + \frac{\delta W[J]}{\delta J(q)} \frac{\delta
  W[J]}{\delta J(-p)} \rb_{q=p} = 0\,.
\end{eqnarray}

It is now easy to rewrite this in terms of $\Gamma$; we substitute
\begin{equation}
J(p) = R(p) \Phi (p) - \frac{\delta \Gamma [\Phi]}{\delta \Phi (-p)}\,,\quad
\frac{\delta W[J]}{\delta J(-p)}= \Phi (p)\,,\quad
\frac{\delta^2 W[J]}{\delta J(p) \delta J(q)} = G_{p,q} [\Phi]
\end{equation}
to obtain (\ref{Kinvariance-Gamma}):
\begin{equation}
- \int_p \frac{\delta \Gamma [\Phi]}{\delta \Phi (p)}
D_\mu^K (p) \Phi (p) + \frac{1}{2}\int_p \left( - p \cdot \partial_p +
  2 - 2 \gamma \right) R(p) \cdot \frac{\partial}{\partial p_\mu}
G_{-p,q} [\Phi]\Big|_{q=p} = 0\,.
\end{equation}

\section{Rewriting $\Sigma_\mu^K = 0$ for $S$\label{appendix-SigmaK}}

In sect.~\ref{section-conformalinvariance}  we have
written the invariance of the Wilson action $S$ under the special
conformal transformation as
\begin{equation}
\Sigma_\mu^K \equiv - e^{-S} \int_p K(p) \frac{\delta}{\delta \phi
  (p)} \left( D_\mu^K (p) \Phi (p) e^S \right) = 0\,,
\end{equation}
where $\Phi (p)$ is given by (\ref{Phi-def}):
\begin{equation}
\Phi (p) \equiv \frac{1}{K(p)} \left( \phi (p) + \frac{k(p)}{p^2}
  \frac{\delta S}{\delta \phi (-p)} \right)\,.
\end{equation}
We wish to rewrite the invariance more explicitly in terms of $S$ and
$\phi$.  Expanding $\Sigma_\mu^K$, we obtain
\begin{eqnarray}
- \Sigma_\mu^K &=& \int_p K(p) \left( D_\mu^K (p) \frac{\delta \Phi
                 (p)}{\delta \phi (q)}\Big|_{q=p} + D_\mu^K (p) \Phi
                 (p) \cdot \frac{\delta S}{\delta \phi (p)} \right)\nn\\
&=& \int_p K(p) \left[ D_\mu^K (p) \left( \frac{k(p)}{p^2 K(p)}
    \frac{\delta^2 S}{\delta \phi (-p) \delta \phi
    (q)}\right)_{q=p}\right.\nn\\
&&\left.\quad + \lb D_\mu^K (p) \left(\frac{1}{K(p)} \phi (p)\right) +
   D_\mu^K (p) \left( \frac{k(p)}{p^2 K(p)} \frac{\delta S}{\delta
   \phi (-p)} \right) \rb \frac{\delta S}{\delta \phi (p)} \right] \,,
\end{eqnarray}
where we have dropped the field independent part.  Using $R(p) =
\frac{p^2 K(p)^2}{k(p)}$, we rewrite this as
\begin{eqnarray}
- \Sigma_\mu^K &=& \int K(p) D_\mu^K (p) \left(\frac{1}{K(p)} \phi
                   (p)\right) \frac{\delta S}{\delta \phi (p)} \nn\\
&& \, + \int_p R (p) D_\mu^K (p) \lb
\frac{K(p)}{R(p)} \left( \frac{\delta^2 S}{\delta \phi (-p) \delta
   \phi (q)} + \frac{\delta S}{\delta \phi (-p)} \frac{\delta
   S}{\delta \phi (q)} \right) \frac{K(q)}{R(q)} \rb_{q=p} \,.
\end{eqnarray}
We can expand
\begin{equation}
K(p) D_\mu^K (p) \frac{\phi (p)}{K(p)}
= D_\mu^K (p) \phi (p) + p \cdot \partial_p \ln K(p) \cdot \frac{\partial \phi
  (p)}{\partial p_\mu} + K(p) D_\mu^K (p) \frac{1}{K(p)} \cdot \phi
(p) \,.
\end{equation}
Using (\ref{formula}), we can rewrite the second integral of
$-\Sigma_\mu^K$ as
\begin{eqnarray}
&&\frac{1}{2} \int_p \left( p \cdot \partial_p - 2 + 2 \gamma \right) R(p) \cdot
\frac{\partial}{\partial p_\mu} \lb
\frac{K(p)}{R(p)} \left( \frac{\delta^2 S}{\delta \phi (-p) \delta
   \phi (q)} + \frac{\delta S}{\delta \phi (-p)} \frac{\delta
   S}{\delta \phi (q)} \right) \frac{K(q)}{R(q)} \rb_{q=p}\nn\\
&&= \frac{1}{2} \int_p \left( p \cdot \partial_p - 2 + 2 \gamma
\right) R(p) \cdot 
\left(\frac{K(p)}{R(p)}\right)^2  \frac{\partial}{\partial p_\mu}
\lb  \frac{\delta^2 S}{\delta \phi (-p) \delta
   \phi (q)} + \frac{\delta S}{\delta \phi (-p)} \frac{\delta
   S}{\delta \phi (q)} \rb_{q=p}\,.
\end{eqnarray}
(Note $K(p)$ and $R(p)$ depend only on $p^2$.)  Hence, we can rewrite
$\Sigma_\mu^K = 0$ as
\begin{eqnarray}
&&\int_p \left( D_\mu^K (p) \phi (p) + p \cdot \partial_p \ln K(p) \cdot
   \frac{\partial \phi (p)}{\partial p_\mu} + K(p) D_\mu^K (p) \frac{1}{K(p)}
   \cdot \phi (p) \right) \frac{\delta
   S[\phi]}{\delta \phi (p)}\nn\\
&& + \frac{1}{2} \int_p \frac{k(p)}{p^2} \left( p \cdot \partial_p \ln
   R (p) - 2 + 2 \gamma \right) \frac{\partial}{\partial p_\mu} \lb
\frac{\delta^2 S}{\delta \phi (-p) \delta \phi (q)} + \frac{\delta
   S}{\delta \phi (-p)} \frac{\delta S}{\delta \phi (q)} \rb_{q=p} =
   0\,.
\label{Kinvariance-S}
\end{eqnarray}
This corresponds to (2.79b) of \cite{Rosten:2016nmc} which differs
slightly from (\ref{Kinvariance-S}) due to a difference in the choice
of cutoff functions.  The similar difference between (2.79a) of
\cite{Rosten:2016nmc} and our ERG differential equation
(\ref{ERGequation}) has been explained in Appendix C of
\cite{Igarashi:2016qdr}.

\begin{acknowledgments}
  The author thanks Prof.~Bala Sathiapalan for encouragement and many
  discussions, and Dr.~Carlo Pagani for his interest in this work.
\end{acknowledgments}

\bibliography{paper}

\end{document}